\documentstyle[12pt]{article}
\newcommand{\gsim}{\mathrel{\hbox{\rlap{\lower.55ex \hbox {$\sim$}}
                   \kern-.3em \raise.4ex \hbox{$>$}}}}
\newcommand{\lsim}{\mathrel{\hbox{\rlap{\lower.55ex \hbox {$\sim$}}
                   \kern-.3em \raise.4ex \hbox{$<$}}}}
\newcommand {\vectwo}[2] {\left(\begin{array}{c}#1\\#2\end{array}\right)}

\newcommand {\beq}    {\begin{equation}}
\newcommand {\eeq}    {\end{equation}}

\newcommand {\thetaw} {\theta_{w}}
\newcommand {\ES}     {E_{{\rm S}}}

\newcommand {\twoi}   {2\, i}

\newcommand {\fZ}     {f_{Z}}
\newcommand {\hZ}     {h_{Z}}

\newcommand {\eZ}     {\epsilon_{Z}}
\newcommand {\stwNCS} {\sin^{2}\theta_{w}^{\rm NCS}}

\newcommand {\nonpert}{non-perturbative}
\newcommand {\ncl} {non-contractible loop}

\newcommand {\ncs} {non-contractible sphere}

\newcommand {\sph}   {spha\-le\-ron}

\newcommand {\cs} {con\-fi\-gu\-ra\-tion spa\-ce}
\newcommand {\ew} {electroweak}

\newcommand {\Sstar} {S$^{\star }$}
\newcommand {\Istar} {I$^{\star }$}

\newcommand {\ewsm}     {electroweak standard model}

\begin{document}
\begin{titlepage}
\hspace*{\fill}
NIKHEF--H / 94-19
\newline
\hspace*{\fill}
June, 1994 
\renewcommand{\thefootnote}{\fnsymbol{footnote}}
\begin{center}
	\vspace{4\baselineskip}
	{\large \bf ANOTHER LOOK AT THE ELECTROWEAK VORTEX SOLUTION }
\footnote{\normalsize
to appear in J. C. Rom\~{a}o (ed.), Electroweak Physics and the Early Universe,
NATO Advanced Research Workshop, Sintra, Portugal, 23--25 March, 1994
         } \\
\setcounter{footnote}{0}
	\vspace{2\baselineskip}
	{\large
	 F. R. Klinkhamer \\
	\vspace{1\baselineskip}
	 CHEAF / NIKHEF--H \\ Postbus 41882 \\
	 1009 DB Amsterdam \\ The Netherlands
	}\\
	\vspace{3\baselineskip}
	{\bf Abstract} \\
\end{center}
We discuss the position in configuration space
of the electroweak vortex solution.
\end{titlepage}
\newpage
\vspace*{10\baselineskip}
\noindent {\bf ANOTHER LOOK AT THE ELECTROWEAK VORTEX SOLUTION}
\vspace{3\baselineskip} \newline
\hspace*{1in} F. R. Klinkhamer \vspace{1\baselineskip}\newline
\hspace*{1in} CHEAF / NIKHEF--H \newline
\hspace*{1in} Postbus 41882 \newline
\hspace*{1in} NL--1009 DB Amsterdam
\vspace{3\baselineskip}

Many, if not most, of the topics discussed at this workshop are
\nonpert ~in origin. For this reason it may be of interest to
report here on one minor, but rigorous, \nonpert ~result obtained
in collaboration with P. Olesen \cite{KO94}.
Similar ideas have been presented at this meeting by M. James.
The goal of this talk then is to suggest a new point of view
for an old classical solution, namely the well-known vortex solution
\cite{NO73} as embedded in the \ewsm ~\cite{N77}.
Specifically, we would like to know where in \cs ~this solution fits in.

Let us start by reviewing what limited knowledge we have of \cs, i. e. the
abstract space of static, finite energy field configurations, with
the gauge freedom eliminated. For simplicity, we consider only the
bosonic fields of the \ewsm ~(the response of the fermionic fields is
certainly important for the physics applications, but the crucial dynamics is
beleived to be carried by the bosonic fields).
The classical vacuum, with vanishing gauge fields and
a constant Higgs field (energy $E_{\rm ~V} = 0$), corresponds to a
single point V in \cs. The energy surface at this point V is
stationary, in other words the vacuum configuration solves the classical
field equations. It turns out that the topology of \cs ~is highly
non-trivial and this leads to the existence of other stationary points,
i. e.  new classical solutions. We mention two of them.

First, there exists a \ncl ~(NCL) in \cs, pa\-ra\-me\-tri\-zed by
$\mu \in [-\pi,+\pi]$, which for $\mu = \pm \pi$ passes through the vacuum V
and for $\mu = 0$ through a new classical solution, the \sph ~S \cite{KM84}.
This NCL captures some of the topology of \cs, because it is based on
a non-trivial map
${\rm S}_{1} \times {\rm S}_{2} \rightarrow {\rm SU(2)} \sim {\rm S}_{3}$,
where ${\rm S}_{1}$ refers to the loop parameter $\mu$,
${\rm S}_{2}$ to the angles on the sphere at spatial infinity and
SU(2) to the non-abelian gauge group of the \ewsm.
In short, the NCL wraps around a ``hole'' in \cs.
The energy of the corresponding \sph ~S is a function of the
mass ratios $M_{H} / M_{W}$ and $M_{Z} / M_{W}$,
its order of magnitude being $\ES = {\rm O}(M_{W}/\alpha) \sim 10\; {\rm TeV}$.
Second, there exists a \ncs ~(NCS) in \cs, parametrized by
$\mu, \nu \in [-\pi,+\pi]$, which for $\mu = \pm \pi$ or $\nu = \pm \pi$
passes through V and for $\mu=\nu= 0$ through another \sph ~\Sstar
{}~\cite{K93b}.
The essential non-trivial map is now
${\rm S}_{2} \times {\rm S}_{2} \rightarrow {\rm SU(2)}$,
where the first ${\rm S}_{2}$ refers to the sphere with
parameters $\mu$ and $\nu$ and the second to sphere at spatial infinity.
The energy of the \sph ~\Sstar ~is a little less than twice that of the
\sph ~S.  
As to the physics applications, the crucial observation is that both
the NCL and the NCS are related to anomalies, respectively the
chiral U(1) anomaly and the global SU(2) anomaly. More concretely,
the \sph ~S seems to play a role in B+L violating processes at
high temperatures \cite{KM84,AM87,Detal}, whereas the \sph ~\Sstar, or
rather its related constrained instanton \Istar, may have to do with the
asymptotics of perturbation theory \cite{K93b,K93a,RS94}.

These considerations were for 3-dimensional configurations of finite energy
$E$, but we can also
restrict ourselves to field configurations which are constant in, say,
the $x_{3}$ direction.  [ Cylindrical coordinates may be defined in
terms of the cartesian coordinates by
$ (\rho \cos \phi\,,  \rho \sin \phi\,, z) \equiv (x_{1},x_{2},x_{3}) $. ]
The dimensionless ``energy'' for these 2-dimensional configurations is then
given by
\begin{eqnarray}
\epsilon &\equiv & \frac{1}{\pi v^{2}} \int dx_{1} dx_{2} \: e \nonumber \\
         &=      & \frac{1}{\pi v^{2}} \int dx_{1} dx_{2}
        \left[ \frac{1}{4 \,g^{2}}\, \left( W_{m n}^{a} \right)^{2} +
        \frac{1}{4 \,g^{\prime \,2}}\, \left( B_{m n}     \right)^{2} +
        |D_{m} \Phi|^{2} +
        \lambda \left( |\Phi|^{2} - v^{2}/2 \right)^{2} \, \right] ,
\label{eq:E}
\end{eqnarray}
with field strengths and covariant derivatives
\begin{eqnarray*}
W_{mn} &\equiv& W_{mn}^{a}\: \tau^{a}  \equiv
  \partial _{m} W_{n} - \partial _{n} W_{m} + [W_{m},W_{n}]\\
B_{m n} &\equiv&  \partial _{m} B_{n} - \partial _{n} B_{m}\\
D_{m}\Phi&\equiv&\left( \partial_{m} + B_{m}/(\twoi) + W_{m} \right) \Phi \\
W_{m} &\equiv& W_{m}^{a}\: \tau^{a} \\
\tau^{a}&\equiv&\sigma^{a}/(\twoi)  \: .
\label{eq:definitions}
\end{eqnarray*}
Here $W$ and $B$ are the SU(2) and U(1) gauge fields and $\Phi$ is the
complex doublet Higgs field.
The semiclassical masses of the $W^{\pm}$ and $Z^{0}$ vector bosons are
$M_{W}=\frac{1}{2}\, g\, v$
and $M_{Z} = M_{W} / \cos \thetaw$, with the weak mixing angle
$\thetaw$ defined as $\tan \thetaw \equiv g' / g$, and the mass
of the physical Higgs scalar is $M_{H}=\sqrt{8\,\lambda/g^{2}}\, M_{W}$.
Our goal now is to investigate the topology of this different \cs, i. e.
the space of finite $\epsilon$ field configurations.
Again, we will do this by constructing a \ncs.

Introducing for the parameters $\mu, \nu \in [-\pi,+\pi]$
the notation $[\mu\nu] \equiv {\rm max}(|\mu|,|\nu|)$,
our NCS of 2-dimensional configurations is given by
\begin{eqnarray}
\pi/2 \leq [\mu\nu] \leq \pi & :
                         &W   =  0  \nonumber\\
                         &&B  =  0  \nonumber\\
                         &&\Phi=  \left( 1 - (1-h) \sin [\mu\nu]\,\right)
                        \: \frac{v}{\sqrt{2}} \: \vectwo{0}{1}  \nonumber\\
                         &&  \nonumber\\
0 \leq[\mu\nu] <   \pi/2 & :
                         &W   = -f\:G^{a} \, \tau^{a}
                                  \nonumber\\
                         &&B  =  f\: \sin^{2}\thetaw \: F^{3}  \nonumber\\
                         &&\Phi=h\:\frac{v}{\sqrt{2}}\:U \vectwo{0}{1},
\label{eq:ZNCS}
\end{eqnarray}
with the following Lie algebra valued 1-forms
\begin{eqnarray*}
F^{a} \, \tau^{a} & \equiv & U^{-1} \, {\rm d}U \nonumber \\
G^{a} \, \tau^{a} & \equiv & U \: \left[
F^{1} \, \tau^{1} +  F^{2} \,\tau^{2} + \cos^{2}\thetaw \: F^{3} \,\tau^{3}
                             \right] \: U^{-1} \: ,
\label{eq:FGdefinitions}
\end{eqnarray*}
$SU(2)$ matrices
\begin{eqnarray*}
M(\mu,\nu,\phi) &=&
\left( \begin{array}{l}
       \:\sin \mu                    \\
       \:\cos \mu \:\sin \nu           \\
       \:\cos \mu \:\cos \nu \:\sin \phi \\
       \:\cos \mu \:\cos \nu \:\cos \phi
       \end{array}
\right)
\cdot
\left( \begin{array}{c}
       -i \sigma_{1} \\
       -i \sigma_{2} \\
       -i \sigma_{3} \\
        1
       \end{array}
\right)\\
\label{eq:M}
&& \nonumber \\
U(\mu,\nu,\phi) &=& M(\mu,\nu,0)^{-1} \; M(\mu,\nu,\phi)
\label{eq:U}
\end{eqnarray*}
and boundary conditions for the axial functions $f(\rho)$ and $h(\rho)$
\begin{eqnarray*}
f(0)=h(0)                           &=& 0  \nonumber\\
\lim_{ \rho \rightarrow \infty} f,h &=& 1  \: .
\label{eq:Zbcs}
\end{eqnarray*}
The topology of \cs ~is captured by the non-trivial map $M$ :
${\rm S}_{2} \times {\rm S}_{1} \rightarrow {\rm SU(2)} $,
where ${\rm S}_{2}$ refers to the sphere with parameters $\mu, \nu$ and
${\rm S}_{1}$ to the circle at spatial infinity.
This map with winding number $m=1$ can be generalized by replacing
$\sin\phi$ and $\cos\phi$ by $\sin m\phi$ and $\cos m\phi$.
Also note that there is no longer the possibility of having a \ncl,
since the map ${\rm S}_{1} \times {\rm S}_{1} \rightarrow {\rm SU(2)} $
is always trivial.

With these configurations the resulting dimensionless energy density is
for $[\mu\nu] < \pi/2$
\begin{eqnarray}
e &=& \cos^{2}\mu \: \cos^{2}\nu \:
      \left(\: 1 -  \cos^{2}\mu \: \cos^{2}\nu  \: \sin^{2}\thetaw\:\right)\:
      \cos^{-2}\thetaw\:\left[ \:\rho^{-2}\:(\partial_{\rho}\,f)^{2}\:\right]
      + \nonumber\nonumber \\
  & & \cos^{2}\mu \: \cos^{2}\nu \;
      \left[ \: \rho^{-2}\:h^{2}\: \left( 1-f \right)^{2} \: \right] +
      (\partial_{\rho} \,h)^{2} +
      \frac{1}{4} \left( \frac{M_{ H}}{M_{ Z}} \right)^{2}
      \left( h^{2} -1 \right)^{2}
\label{eq:Zeballoon}
\end{eqnarray}
and for $\pi/2 \leq [\mu\nu] \leq \pi$
\begin{eqnarray}
e &=&
    (\partial_{\rho} \,k)^{2} +
    \frac{1}{4} \left( \frac{M_{ H}}{M_{ Z}} \right)^{2}
    \left( k^{2} -1 \right)^{2} \: ,
\label{eq:Zecord}
\end{eqnarray}
with $k \equiv 1 - (1-h) \sin [\mu\nu] $ .
Perhaps the main subtlety in the ansatz (\ref{eq:ZNCS}) is the way the
SU(2) and U(1) gauge fields are distributed,
in order to keep their total kinetic energy density (\ref{eq:Zeballoon})
down by partially cancelling the $\cos^{-2}\thetaw$ factor.

We now observe that the NCS configuration  at $\mu = \nu =0$
gives precisely the \ew ~vortex solution ($Z$-string).
Just recall that the fields of the $Z$-string
can be written in the following form \cite{NO73,N77}
\begin{eqnarray}
W    &=& - \cos^{2}\thetaw \: \fZ \: {\rm d}V\: V^{-1} \nonumber\\
B    &=& - \tan^{2}\thetaw \: W^{3}  \nonumber\\
\Phi &=& \hZ \:\frac{v}{\sqrt{2}}\:V \vectwo{0}{1},
\label{eq:NO}
\end{eqnarray}
where the $SU(2)$ matrix $V$ is simply
\[
V =  \cos \phi \: 1 - \sin \phi \: i \sigma_{3} \equiv U(0,0,\phi)
\]
and the functions $\fZ$ and $\hZ$ are the solutions of the
field equations with the standard boundary conditions.
The corresponding energy density of the $Z$-string
\beq
e_{Z} =
      \rho^{-2}\:(\partial_{\rho} \, \fZ)^{2} +
      \rho^{-2}\:\hZ^{2}\: \left( 1-\fZ \right)^{2}+
      (\partial_{\rho} \, \hZ)^{2} +
      \frac{1}{4} \left( \frac{M_{ H}}{M_{ Z}} \right)^{2}
      \left( \hZ^{2} -1 \right)^{2}
\label{eq:eNO}
\eeq
is independent of the value of the weak mixing angle $\thetaw$.

With the axial functions of the $Z$-string $f = \fZ $ and $h = \hZ$,
the energy  profile over the NCS (\ref{eq:ZNCS})
is for $0 \leq[\mu\nu] < \pi/2$
\begin{eqnarray}
\epsilon(\mu,\nu) &=& \epsilon_{Z} \left[ \;
 \cos^{2}\mu \: \cos^{2}\nu   \left(
 \frac{ 1 -  \cos^{2}\mu \: \cos^{2}\nu  \: \sin^{2}\thetaw} {\cos^{2}\thetaw}
 \; a + b \right) + c + d \; \right] \\
\label{eq:epsilonZNCSglobal}
&& \nonumber\\
  &=& \epsilon_{Z} \left[\; 1 + (\mu^{2} + \nu^{2})
      \left( \tan^{2}\thetaw - 1 - b / a \right) a +
      {\rm O}(\mu^{4},\, \nu^{4},\, \mu^{2}\nu^{2})\; \right] \;,
\label{eq:epsilonZNCSlocal}
\end{eqnarray}
in terms of the following integrals
\begin{eqnarray*}
a  &\equiv& \epsilon_{Z}^{-1} \:\int_{0}^{\infty} d\rho\:
            \rho^{-1}\:\left(\partial_{\rho} \, \fZ\right)^{2}  \nonumber \\
b  &\equiv& \epsilon_{Z}^{-1} \;\int_{0}^{\infty} d\rho\:
            \rho^{-1}\:\hZ^{2}\: \left( 1-\fZ \right)^{2} \nonumber \\
c  &\equiv& \epsilon_{Z}^{-1} \;\int_{0}^{\infty} d\rho\, \rho \:
            \left(\partial_{\rho} \, \hZ\right)^{2} \nonumber \\
d  &\equiv& \epsilon_{Z}^{-1} \;\int_{0}^{\infty} d\rho\, \rho \;
            1 / 4 \, \left( M_{ H}/ M_{ Z} \right)^{2}
            \left( \hZ^{2} -1 \right)^{2}  \; ,
\label{eq:abcd}
\end{eqnarray*}
which add up to 1.
{}From (\ref{eq:epsilonZNCSlocal}) there is
manifest instability of the $Z$-string for $ \thetaw \leq \pi/4$
and arbitrary Higgs mass.
In short, the $Z$-string of the \ew ~interactions (with the experimental
value $\thetaw \sim \pi/6$)
is the 2-di\-men\-si\-o\-nal spha\-le\-ron of a \ncs.

Numerical results allow
us to extend the range of instability beyond $\sin^{2}\thetaw = 1/2$.
In Table 1 we give the maximum allowed value $\thetaw$ for instability
according to (\ref{eq:epsilonZNCSlocal}).
The $Z$-string  solution has for $\thetaw < \thetaw^{\rm NCS}$ two
negative modes, made explicit by the fields (\ref{eq:ZNCS}) of the NCS.
This range of instability is consistent with the results of \cite{JPV93}.
\begin{table}[t]
\begin{center}
\begin{tabular}{lccccccc}
\hline
\hline
           & \multicolumn{7}{c}{$M_{H}/M_{Z}$}    \\
           &$1/8$  &$1/4$  &$1/2$  &$1$    &$2$    &$4$    &$8$   \\
\hline
$\eZ$      &$0.47$ &$0.59$ &$0.76$ &$1.00$ &$1.34$ &$1.79$ &$2.35$ \\
$a$        &$0.16$ &$0.18$ &$0.20$ &$0.21$ &$0.21$ &$0.20$ &$0.18$ \\
$b$        &$0.19$ &$0.22$ &$0.25$ &$0.29$ &$0.35$ &$0.43$ &$0.52$ \\
$c$        &$0.50$ &$0.43$ &$0.36$ &$0.29$ &$0.22$ &$0.16$ &$0.12$ \\
$d$        &$0.15$ &$0.17$ &$0.19$ &$0.21$ &$0.22$ &$0.21$ &$0.18$ \\
$\stwNCS$  &$0.69$ &$0.69$ &$0.69$ &$0.70$ &$0.73$ &$0.76$ &$0.79$ \\
\hline
\hline
\end{tabular}
\end{center}
\caption[]{ \protect \small
Numerical results for the $Z$-string solution :
the dimensionless energy $\eZ$, the integrals $a$--$d$ and the
critical mixing angle $\theta_{w}^{\rm NCS}$.
}
\end{table}

These numerical results give also the complete energy profile (7)
of the non-con\-trac\-ti\-ble sphere through
the $Z$-string, as a function of both $\thetaw$ and $M_{H}/M_{Z}$.
For the case of perturbative stability \cite{JPV93} of the $Z$-string,
we find an energy barrier for decay towards the vacuum.
Of course, this is not yet the minimal energy barrier, but it may be
indicative.
Taking the parameter values $M_{H}/M_{Z} = 1/8$ and $\sin^{2}\thetaw = 63/64$,
for example, we easily find the
$\mu = 0$ section of the energy profile $\epsilon(\mu,\nu)$ over the NCS :
starting at $\nu=0$, with $\epsilon=\eZ = 0.47$, the energy rises
to a maximum value $\epsilon = 1.57$ at $\nu \sim \pi/4$
and then drops to zero as $\nu \rightarrow \pi$.

To conclude, we have obtained a topological understanding for the
existence and generic instability of the \ew ~vortex solution.
\vspace{1\baselineskip} \newline
\noindent {\Large \bf Acknowledgements}
\vspace{1\baselineskip} \newline
It is a pleasure to thank F. Freire, J. C. Rom\~{a}o and the other
organizers for bringing the Early Universe (and us) to Sintra.
\vspace{1\baselineskip} \newline

\end{document}